\documentclass[10pt,twoside]{article}
\usepackage[applemac]{inputenc}
\usepackage{amssymb,amsmath}
\def\volnum{Manuscrit, Revision 4, 15 Septembre 2006 }

\usepackage{aflbcours}
\pdebut{1}
\def\pacs#1{\LP P.A.C.S.: #1}
\title{On solutions of the standard-model Lagrangian with
a Majorana mass term}
\titleshort{Standard model with Majorana mass}
\author{Rainer Plaga}
\authorshort{R. Plaga}
\address{Franzstr. 40
        \\D-53111 Bonn, Germany}

\begin{document}
\maketitle

\vskip 1cm
\begin{abstract}
{\it 
ABSTRACT. It is demonstrated that the standard-model Lagrangian with 
a Majorana mass term for the neutrino admits no non-trivial solution in the presence of
charged leptons. 
Because 
the standard model is generally believed to describe the gauge interactions of neutrinos 
correctly,
the Majorana mass term must vanish and thus cannot enable neutrino-less double $\beta$ decay.
More generally, neutrinos with standard-model gauge interactions
cannot be Majorana fields.
Historical reasons why this conclusion has not been drawn earlier are analyzed.}  
\end{abstract}

\section{Introduction}
\subsection{Majorana's lesson and definition of his field}
Ettore Majorana  
presented a crucial insight with eq.(10) of his ultimate publication ``Teoria simmetrica 
dell'elletrone e del positrone''\cite{majorana}.
In a suitable
representation of the $\gamma$ matrices\footnote{Nowadays called
``Majorana representation''.} 
the real part of a spinor alone is
a solution to the free Dirac equation.
This can be expressed in a representation independent
way as: spinor operators ``$\Psi$'' that are self-charge conjugate, i.e. 
for which: 
\begin{equation}
\Psi = \Psi^c \ \ \ {\rm (Majorana \ \ condition)}
\label{mc}
\end{equation}
fulfill the free Dirac equation.
Here the superscript ``$^c$'' symbolizes the operation 
of ``charge conjugation''.
With our notation\footnote
{Appendix \ref{app2} contains the conventions I chose.}
charge conjugation is complex conjugation in the
Majorana representation:
\begin{equation}
\Psi = \Psi^c = \Psi^* \ \ \ {\rm (Majorana \ \ condition \ \ and \ \ representation )}
\label{mc2}
\end{equation}
The operation takes a slightly
different form in other representations (see appendix \ref{app2}).
Self charge-conjugate neutrinos (i.e. neutrino fields that fulfill
condition \ref{mc})
are called ``Majorana neutrinos''\footnote{In a recent review\cite{bile} my eq.(\ref{mc})
appears as eq.(8) and my eq.(\ref{mc2}) as a unnumbered eq. a few lines below eq.(8).}. 
\subsection{Aim and basic assumptions of this manuscript}
\label{aim}
Majorana taught us that the equation of motion (or the Lagrangian
from which it can be derived) is primary. Majorana fields are one
special solution of the free Lagrangian. Whether such a solution is still allowed or even required
when further mass and/or interaction terms are added to the free Lagrangian
is a non-trivial and fundamental question. 
Surprisingly it has not been comprehensively addressed up to today.
In this manuscript
I answer it for an important special case: 
the inclusion of standard-model interactions and a (non standard-model) Majorana
mass term.
\\
In this manuscript I assume that the usual Dirac equation
with standard-model gauge interactions
is valid. It has been suggested that Majorana neutrinos
exist in theories beyond this framework
(e.g. \cite{dvoe}).
This important work is beyond the scope of the present paper.
\subsection{Structure of this manuscript}
The rest of the introduction (subsection \ref{freeL}) reviews Majorana mass terms.
Section \ref{tot} analyzes the solution of a neutrino Lagrangian 
with standard-model interactions and the addition of a (non standard-model) Majorana mass term.
Section \ref{why} identifies historical stumbling blocks in the way
of an earlier understanding of the problem addressed here. 
Section \ref{concl} concludes. Appendix \ref{app2} lists notational
conventions, appendix \ref{app}) reviews 
chiral spinors for the convenience of the reader, appendix \ref{app3}
proves that Majorana masses require Majorana solutions and appendix \ref{free}
reviews the equivalence of free Majorana and Weyl neutrinos.

\subsection{Solutions of a free Lagrangian with a Majorana mass term}
\label{freeL}
The following Lagrangian for a free neutrino 
``requires'' a Majorana
solution because it contains a ``Majorana mass term''\footnote{Majorana\cite{majorana} used only Dirac mass
terms that allow but do 
not require Majorana solutions. 
The Majorana nature of a Lagrangian's mass term is a sufficient but not necessary condition for 
the existence of a Majorana-field solution.
Majorana-mass terms were introduced by Jehle\cite{jehle} and its relation
to Majorana fermions was clarified by Serpe\cite{serpe}.}:
\begin{equation}
L_{\nu}^{\rm free}= 
i \bar{\nu} \gamma^{\mu} {\partial \over {\partial x^\mu}} \nu
- \left[ m_{\rm maj} {\bar{\nu}} (\nu)^c/2 + H.C. \right]
\label{majm}
\end{equation}
```$\nu$'' stands for the neutrino-field operator
and $m_{\rm maj}$ is the Majorana mass\footnote{In a recent review\cite{bile} this Majorana
mass term appears as eq.(26)}.
An equation of motion for a 
free neutrino with a Majorana mass - derived from eq.(\ref{majm})
according to the Euler-Lagrange 
equations is:
\begin{equation}
i \gamma^{\mu} {\partial \over {\partial x^\mu}} \nu
- m_{\rm maj} \nu^c = 0 \ \ \ {\rm (Majorana \ \ equation)}
\label{majo}
\end{equation}
The second term contains the charge conjugate of the
field that appears in the first term. 
Such an equation can only be solved in a non trivial way by
inserting a self charge-conjugate field\footnote{
That the Majorana mass term determines the
Majorana nature of the field seems to be
uncontroversial. Bilenky writes\cite{bile}:``(The) nature
of neutrinos ... is determined by the type of mass term.''}. 
Otherwise the terms cannot cancel for non-vanishing neutrino fields.
For a detailed proof see appendix \ref{app3}.
Because particles cannot be turned to their charge conjugate
by any Lorentz boost, this conclusion holds for any Lorentz
frame\footnote{In the words of Pauli\cite{pauli}:`` The ordering between particles
and antiparticle solutions is Lorentz invariant.''}. I summarize this in the following 
\\
{\bf Theorem A}
\\
{\it If the Lagrangian contains a finite Majorana-mass term (defined in eq.(\ref{majm})), then
all solutions in all inertial frames are Majorana fields, 
i.e. they must fulfill eq.(\ref{mc}).}
\section{Solutions of the Standard Model Lagrangian with a Majorana mass term}
\label{tot}
My aim is to answer the following question: What are the solutions
to the full Lagrangian valid for a physical neutrino with Majorana-mass and interaction terms? 
\subsection{Form of the total Lagrangian}
While it can never be excluded that
``new physics'' will modify a theory, it is generally believed that SM\cite{iz}
{\it interactions} will always at least remain a good approximation to a ``final theory''
for the conditions prevalent in current experiments\footnote{Bilenky\cite{bile} assumes a
SM interaction term in his review of Majorana neutrinos because it
``perfectly describes existing weak interaction data.''}.  
The situation is of course different for the mass term: here it is
currently believed that it might be a Majorana mass term. Such a term
is qualitatively different from
the usual SM Higgs terms that confer mass to all non-neutrino fermions.
\\
The total neutrino Lagrangian with 
a SM interaction and Majorana mass term is:
\begin{eqnarray}
L_{\nu}^{\rm tot} =
i \bar{\nu} \gamma^{\mu} {\partial \over {\partial x^\mu}} \nu
+  {g \over 2\sqrt{2}} \left[W_{\mu}^+ \bar{\nu} \gamma^{\mu} (1 - \gamma_5) e^- + H.C. \right]
\nonumber
\\
- \left[ m_{\rm maj} {\bar{\nu}} (\nu)^c/2 + H.C. \right]
\label{sml}
\end{eqnarray}
Here $e^{-}$ symbolizes the electron (or muon/tau) field.
$g$ is a gauge coupling constant and W$^+$ is the massive charged
weak boson field.
For brevity neutral-current terms have been omitted. Their inclusion
would not change any conclusions of this manuscript.
\\
The full Majorana equation 
for the neutrino field $\nu$ with charged-current interaction derived from eq.(\ref{sml}) is:
\begin{equation}
i \gamma^{\mu} {\partial \over {\partial x^\mu}} \nu
+ {g \over 2\sqrt{2}} \left[W_{\mu}^+ \gamma^{\mu} (1 - \gamma_5) e^- \right] 
- m_{\rm maj} \nu^c = 0
\label{full}
\end{equation}
\subsection{Ultra-relativistic solutions of the SM Lagrangian}
\label{ultra}
In the ultra-relativistic limit m/E $\rightarrow$ 0 the mass term is finite
but negligible compared to the kinetic term\footnote{A Lorentz frame in which the direction of motion
is the same as it would be for a vanishing rest mass is assumed.} and we
set it to 0 here.
\\
At this point I recommend to consult the appendix \ref{app}, 
to review the definition of ``left/right handed states'' used here. {\bf
Throughout this manuscript they
denominate the chirality and {\it not} the helicity eigenstates.}
\\
Factoring out
$\gamma^{\mu}$ in eq.(\ref{full}) and writing the fields as
column vectors in the chiral representation leads to:
\begin{equation}
i \gamma^{\mu} \left[ {\partial \over {\partial x^\mu}} \left( \begin{array}{c} \nu_1 \\ \nu_2\end{array} \right)
+  {g \over 2\sqrt{2}} \sqrt{2} W_{\mu}^+ \left( \begin{array}{c} 0 \\ e_2^- \end{array} \right) \right] 
 = 0
\end{equation}
One concludes that the derivative of right handed component of the
kinetic term $\nu_1$ must be zero likewise:
\begin{equation}
{\partial \over {\partial x^\mu}}
\nu_1 =
 {\partial \over {\partial x^\mu}} \nu_R =
 0
\end{equation}
Let us choose $\nu$=0 at a distant spatial boundary
(all observed neutrinos
were ``produced'' by weak interactions).
If $\nu_R$ = 0 at the boundary and the space-time
derivatives are 0 in general $\nu_R$ cannot change
and remains 0 everywhere. If $\nu_R$=0 any solution
must be left-handed everywhere.
I summarize this conclusion
following theorem:
\\
{\bf Theorem B}
\\
{\it Any solution to
eq.(\ref{full}) in the ultra-relativistic limit 
with a non-vanishing electron amplitude is  predominantly left-handed,
i.e. its chirality= -- 1 to good approximation.}

\subsection{There can be no Majorana neutrinos with SM interactions}
\label{ccs}
Solutions of  a Majorana-massive neutrino created by SM interactions 
(i.e. of Lagrangian eq.(\ref{sml})) must fulfill both theorem A
(``they must be self-charge conjugate''\cite{bile}) and B.(``they must be chiral''\cite{iz}).
My novel insight is that they there is no way that they can be
both fulfilled at the same time:
\\
{\it 1. Theorem B (subsection \ref{ultra}) tells us that
all physical ultra-relativistic neutrinos physical are 
left-handed, i.e. they have a definite chirality = --1.}
\\
{\it 2. Theorem A (subsection \ref{freeL}) tells us
that the Majorana equation (eq.(\ref{mc})) must be fulfilled in all inertial
frames. Ultra-relativistic Majorana fields must be self-charge conjugate.}
\\
{\it 3. Charge conjugation flips chirality (see appendix \ref{app}).
Therefore the Majorana condition can only be fulfilled
for fields that have component of left- and right-handed
fields that are equal in amplitude. 
Such fields are no eigenvector of $\gamma_5$, i.e. they
have no definite chirality. From requirement 2. we conclude
that ultrarelativistic Majorana field cannot have a definite chirality.}
\\
Requirements 1. and 3. can obviously not
be simultaneously fulfilled.
\\
This contradiction forbids any
solution for the physical neutrino field (except the trivial solution $\nu$=0).
A SM Lagrangian with
Majorana-mass term in the presence of electrons 
allows no solutions.
The wide spread folklore\footnote{I write ``folklore'' because - as already mentioned below -
the question posed in this manuscript was never raised in a 
precise manner in the literature.
Rather the existence
of a Majorana solution of the SM Lagrangian with
Majorana-mass term is taken for granted without critical discussion.
}
that it allows Majorana-neutrino solutions
is thereby erroneous. 
\\
Even in theories with ``new physics'' the neutrino's gauge interactions
are expected to be described
to good approximation by the SM in processes realized in today's laboratory.
Necessary conclusions are that
physical neutrinos are no Majorana fields to good approximation. Thereby Majorana mass terms
cannot enable neutrino-less double $\beta$ decay with SM
interactions. 
More generally a Lagrangian with SM interaction terms cannot
have any Majorana solution.
\\
Under the assumptions stated in section \ref{aim} the only
remaining possibility is that neutrinos receive their 
experimentally determined masses through the
ordinary Higgs mechnism, like all other fermions, i.e. the neutrino Lagrangian
has a Dirac mass term.
\section{Why was the conclusion of this manuscript not drawn earlier?}
\label{why}
The conclusion of subsection \ref{ccs} (``neutrinos
are no Majorana fields'') could have been drawn since physicists
endorsed the SM gauge i.e.
since about 1979 (the  year with Nobel prices 
for discovering the SM)\footnote{The question: ``Could the conclusion have been
drawn after the formulation of V-A (in 1958)?'' is difficult to answer.
Before the formulation and experimental confirmation of the SM, 
weak interactions were described by a phenomenological nonrenormalizable current-current
theory, valid only at distances large compared to 1/M$_W$. Probably one could have
already guessed at the result of the present manuscript, but for
a firm conclusion the formulation of a consistent physical theory - the SM - was
necessary.}.
Why the long delay? Below I offer some 
explanations.
\subsection{Modifying the SM interaction?}
It has been suggested that
one should postulate a Lagrangian without neutral-current vector terms
in order to allow for Majorana solutions\cite{kaysershrock}.
This proposal 
modifies the SM interactions of the neutrino.
If we want to avoid the exclusion of Majorana neutrinos 
by the argument in subsection \ref{ccs},
we must adopt a purely pseudo-vectorial interaction term also for charged currents.
However, such a source term  
does not violate parity, in flagrant contradiction to experience.
This rules out this proposal, at least in this simple form.
\subsection{The ``practical Majorana-Dirac confusion theorem'' is incorrect}
Perhaps the most important factor eclipsing the realization that
SM interactions forbid Majorana fields is the widely accepted validity
of an erroneous ``practical Majorana-Dirac confusion theorem''\cite{kayser}.
It asserts that in the ultrarelativistic limit there is no phenomenological difference
between Dirac and Majorana neutrinos if they interact only with V-A interactions.
It is widely accepted as correct,
even though - to my knowledge - no complete proof of the theorem was ever claimed.
In particular the present manuscript seems to be the first that 
systematically discusses the effect of SM charged currents on phenomenological equivalence.
\\
In appendix
\ref{free} I review
a ``Majorana-Dirac confusion theorem'' theorem for the {\bf free} case
(i.e. the case of a {\bf non-interacting} neutrino). 
{\bf This} theorem is correct. It was proven in a flurry of papers in the spring of
1957 (e.g. \cite{mclennan,tou}), i.e. before the V-A structure of the weak current was 
fully understood.
Reviewing the historical literature of the late 1950s
to early 1960s (e.g. \cite{ryan}) I suspect that this theorem
was erroneously thought to be valid in general (i.e.
even if the neutrino has arbitrary interactions) by many. This 
over-interpretation tilled the soil for
the mistaken acceptance of the ``practical Majorana-Dirac confusion theorem'' in 
the early 1980s.
\\
In subsection \ref{wrong}
I demonstrate explicitly
that ultra-relativistic Majorana and Weyl neutrinos have different charge-current interactions 
in the SM. 
Thereby the ``practical Majorana-Dirac confusion theorem'' does not hold
with the V-A interactions prescribed by the SM.

\subsection{The non equivalence of ultra-relativistic Majorana and Weyl neutrinos
in the presence of a V-A interaction term}
\label{wrong}
Formally the Majorana state that is kinematically equivalent to the Weyl neutrino $\nu_L$ is:
\begin{equation}
\nu_M(h=-1) = 1/\sqrt{2}(\nu_L + (\nu_L)^c)
\label{min}
\end{equation}
The first term is a left-handed neutrino amplitdue, that
has helicity = --1.
The second term is a right-handed {\bf anti}neutrino amplitude,
that likewise has helicity = --1
(see appendix \ref{app} for further explanation).
The Majorana state kinematically equivalent to the Weyl antineutrino $(\nu_R)^c$ is:
\begin{equation}
\nu_M(h=+1) = 1/\sqrt{2}((\nu_R)^c + \nu_R)
\label{plu}
\end{equation}
Eqs.(\ref{min},\ref{plu}) are the {\bf only} possibilities to
express Majorana states of definite helicity.
The first term of both states is left-handed and the second is of equal amplitude 
and right handed, i.e. the states conform to requirement 3. 
for Majorana field in section (\ref{cc}). 
\\
One recognizes immediately that 
standard-model (V-A)
interactions do discriminate the states
in eqs.(\ref{min},\ref{plu})
from the Weyl states: both contain an amplitude that
is right handed, therefore sterile 
and cannot be produced by SM charged currents\footnote{Seen
from a different angle: A production
of e.g. state eq.(\ref{min}) together with an electron would not conserve lepton number.
But it is well known that SM interactions do, at least for energies accessible in the
laboratory\cite{dine}.}. 
The equivalence of Dirac and Majorana neutrinos in the
ultra-relativistic limit ceases to hold in the presence of charged-current V-A interactions.
\section{Summary}
\label{concl}
Whether the physical neutrino is described by a Majorana field is determined
solely by the Lagrangian it obeys. 
\\
1. All solutions of the 
standard-model interaction with charged leptons, i.e. including all solutions describing 
charged-current produced neutrinos, must have
a definite chirality (left-handed) in the ultra-relativistic limit.
\\
2. By definition Majorana neutrinos are self-charge conjugate
(Majorana condition eq.(\ref{mc})), a property that holds
in all inertial frames.
\\
3. Because charge conjugation flips chirality, Majorana neutrinos 
must have an equal left-handed and right-handed amplitude
in all inertial frames, i.e. they do not have a definite chirality
because they are no eigen vectors of $\gamma_5$.
\\
Sentence 3. - a corollary from the Majorana condition - 
and the requirements from the SM interactions (sentence 1.) are therefore 
in contradiction and cannot be fulfilled simultaneously.
Because it seems nearly certain that the standard model
describes the weak interactions of neutrinos approximately correctly, neutrinos are 
probably no Majorana fields.
\\
The addition a non-standard model Majorana mass term to the standard-model Lagrangian
enforces a Majorana solution.
Due to the above contradiction the Lagrangian has
no non-trivial solutions describing the production of neutrinos then, i.e. such
a mass term is in disagreement with observation and cannot enable neutrino-less
double $\beta$ decay.
\\
In principle 
neutrinos can have Majorana masses. Neutrinos
interact with SM interactions. 
It was my only aim to show that these
two statements cannot both apply to physical neutrinos.
Because the truth of the latter seems certain, the former must be wrong.

\section{Acknowledgements}
Silvia Pezzoni informed me that I did not understand the first
thing about Majorana neutrinos during a stroll along the Neckar river in 1994.
Alvaro de Rujula constructively criticised draft versions of this
manuscript (in 2006). Both inputs were of crucial importance to this
manuscript. Valeri Dvoeglazov helpfully requested 
clarifications and proofs.  Thanks!
\section{Appendix - notational conventions}
\label{app2}
The notation used is the same as the one used e.g. by Itzykson \& Zuber\cite{iz},
except that a conventional phase factor in the definition
of charge conjugation is chosen as 1 instead of i. It is also the
same as the one used by Landau \& Lifsits\cite{ll} except that they use
a definition of $\gamma_5$ that is different by a factor --1.
All $\gamma$
matrices in explicit notation in the usual representations can be
found in these textbooks.
\\
The components of the metric tensor g$_{\mu\nu}$ are given as
g$_{00}$=+1, g$_{11}$ = g$_{22}$ = g$_{33}$ = --1, all other components
are zero. An arbitrary factor in the definition of charge
conjugation is chosen such that:
\begin{equation}
\Psi^c = \gamma^2 \Psi^*
\label{cc}
\end{equation}
in the standard representation. Because the transformation matrices
from the conventional to the chiral (spinorial) representation
are Hermitian, this relation also holds in this representation.
\\
The chirality operator $\gamma_5$ is defined as:
\begin{equation}
\gamma_5 = i \gamma^0 \gamma^1 \gamma^2 \gamma^3 = \left( \begin{array}{cc}1 &  
0\\ 0 & -1 \\
\end{array} \right) .
\label{g5}
\end{equation}
The matrix is given in the chiral representation.

\section{Appendix - chirality and helicity reviewed}
\label{app} 
The material summarized in this appendix
is explained in more detail e.g. in sections 2-2-1 and 2-4-3
of Itzykson \& Zuber\cite{iz}.
\\
Left- and right-handed fields are defined as eigen vectors
of the operator $\gamma_5$.
``Chirality'' is their respective eigenvalue. 
A field (or antifield) $\Psi_L$ for which $\gamma_5$ $\Psi_L$ = -- $\nu_L$
(eigenvalue --1)
is usually called ``left-handed'' and is labelled with the subscript ``L''. One for which 
$\gamma_5$ $\Psi_R$ = $\Psi_R$ (eigenvalue +1) is called ```right handed'' and
is labelled with the subscript ``R''. The terms ``left- and right-handed'' are {\bf only}
used in this sense in this manuscript and {\bf not} as designating helicity.
In the chiral representation we can then write the spinor as $\Psi$ =
$\left( \begin{array}{c} \phi_1 \\ \phi_2\end{array} \right)$
and $\Psi_L$ = $\left( \begin{array}{c} 0 \\ \phi_2\end{array} \right)$,
$\Psi_R$ = $\left( \begin{array}{c} \phi_1 \\ 0\end{array} \right)$.
\\
The helicity of a spinor is defined as 
the normalized scalar product
between the particle's momentum and its spin vector.
Chirality is not the same as helicity. 
A left-handed particle has chirality=--1 and helicity=--1.
A left-handed {\bf anti}particle has chirality=--1 but its helicity=+1
(\cite{iz} eq.(2-103) ff.)\footnote{There is a nice intuitive explanation of
this fact at the end of section 10.12 of Bjorken/Drell\cite{bjdr}}.
\\
Charge conjugation ``flips chirality'' i.e. it takes left handed particles 
into right handed antiparticles and vice versa. E.g. it
follows from eq.(\ref{cc}): 
\begin{equation}
(\Psi_L)^c = \gamma^2 (\Psi_L)^* = \left( \begin{array}{cc}0 &  
\sigma^2\\ - \sigma^2 & 0 \\
\end{array} \right) 
\left( \begin{array}{c}0 \\ \phi_2^* \end{array} \right) =
\left( \begin{array}{c}\sigma^2 \phi_2^* \\ 0\end{array} \right) =
(\Psi^c)_R.
\end{equation}
Charge conjugation does {\bf not} flip helicity, because a left handed particle
has the same helicity (--1) as a right handed antiparticle.

\section{Appendix - proof that a Majorana mass term requires
a Majorana solution}
\label{app3}
Let us choose the Majorana representation here.
Eq.(\ref{majo}) can be rewritten as:
\begin{equation}
\left[ {i \gamma^{\mu} \over m_{\rm maj}}  {\partial \over {\partial x^\mu}} \right] 
(\nu_{\rm Re} + i \nu_{\rm Im})
- (\nu_{\rm Re} - i \nu_{\rm Im}) = 0 
\label{majo2}
\end{equation}
$\nu_{\rm Re}$ and $\nu_{\rm Im}$ are the real and imaginary amplitude 
of the neutrino field. If the field is complex these (real) amplitudes 
and their spatial and temporal derivatives must
be equal everywhere in space-time.
Because all $\gamma$ matrices are imaginary\cite{iz}, the pre-factor 
${i \gamma^{\mu} \over m_{\rm maj}}$ is
real. 
Eq.(\ref{majo2}) can therefore be written as:
\begin{equation}
(c-1) \nu_{\rm Re} + i (c+1) \nu_{\rm Im} = 0
\label{majo3}
\end{equation}
where c is a real number.
This equation has no solution for equal $\nu_{\rm Re}$
and $\nu_{\rm Im}$. It can only be solved if one of
these amplitudes vanishes, i.e. $\nu$ is either real
or purely imaginaty.
A purely real
field is a Majorana field by eq.(\ref{mc2}).
A purely imaginary field is also a Majorana field, because
in this case $\nu^c$ = -- $\nu$ holds, which differs from 
eq.(\ref{mc}) only by a phase factor.

\section{Appendix - The equivalence of ultra-relativistic {\bf free} Majorana and Weyl
neutrinos}
\label{free}
In the ultra-relativistic limit
m/E $\rightarrow$ 0 {\bf free} Dirac\footnote{Massless Dirac fields
were first studied and recognized as 2-state systems by Weyl\cite{weyl}. ``Weyl
neutrinos'' do not obey the Majorana condition (eq.(\ref{mc})).} and Majorana neutrinos 
are phenomenologically equivalent\cite{mclennan}. 
Dirac neutrinos (with lepton number +1) have helicity h=--1,
Dirac antineutrinos (with lepton number --1) have helicity h=+1.
Helicity is conserved in the ultra-relativistic limit, and it
can therefore in principle assume the role of lepton number for Majorana neutrinos.
\\
However, the validity of this equivalence cannot be simply
taken for granted if the field interacts.
As an obvious counter example: a purely neutral-current vectorial interaction 
(like ``electrical charge'') is possible 
for Dirac but not for
Majorana fields, thus discriminating them, even for m/E $\rightarrow$ 0\footnote{
Scalar (S), pseudo-scalar (iP) and pseudo-vector (A) terms do not change under charge conjugation.
A phenomenological Dirac-Majorana equivalence can only hold for fields with these neutral-current
interactions.}.
\vskip 30pt
\begin{fref}



\bibitem{majorana}
E. Majorana, Nuovo Cimento, Ser.8 {\bf 14}, 171-184, (1937).

\bibitem{bile} S.M. Bilenky, hep-ph/0605172 (2006).

\bibitem{dvoe} V.V. Dvoeglazov, Acta Phys.Polon. {\bf B29}, 619-627, (1998); and
references therein.

\bibitem{jehle} H. Jehle, Phys. Rev.  {\bf 75}, 1609, (1949).

\bibitem{serpe} J. Serpe, Phys. Rev.  {\bf 76}, 1538, (1949).

\bibitem{iz} C. Itzykson, J. Zuber, {\it Quantum Field Theory}, New York, McGraw Hill,
1985.

\bibitem{pauli} W. Pauli, Rev. Mod. Phys. {\bf 13}, 203-232, (1941).

\bibitem{kaysershrock} B. Kayser, R.E. Shrock, Phys. Lett. {\bf 112B}, 137-142, (1982).

\bibitem{kayser} B. Kayser, Phys. Rev. {\bf D 26}, 1662-1670, (1982).

\bibitem{mclennan} J.A. McLennan Jr., Phys. Rev. {\bf 106}, 821-822, (1957).

\bibitem{tou} L.A. Radicati, B. Touschek, Nuovo Cimento {\bf 5},1691-1699, (1957).

\bibitem{ryan} C. Ryan, S. Okubo, Nuovo Cimento Suppl. {\bf 2}, 234-242 (1964);
S. Okubo, priv.comm. (2006).

\bibitem{dine} M.Dine et al., Nucl.Phys. {\bf B342},381-408, (1990).

\bibitem{ll} L.D. Landau, E.M. Lifsits, {\it Teoria Quantistica Relativistica},
Roma, Editori Riuniti, 1978.

\bibitem{bjdr} J.D. Bjorken, S.D. Drell, {\it Relatistic Quantum Mechanics}, New York, McGraw Hill, 1964.

\bibitem{weyl}
H. Weyl, Z.Phys. {\bf 56}, 330-352, (1929).
\end{fref}

\man{15 septembre 2006}
\end{document}